# ANALYSIS OF ASYMMETRIC PIEZOELECTRIC COMPOSITE BEAM

*Jung-San Chen[1]\*, Shyh-Haur Chen[2], and Kuang-Chong Wu[3]*

Institute of Applied Mechanics, National Taiwan University, Taipei 106, Taiwan

## ABSTRACT

This paper deals with the vibration analysis of an asymmetric composite beam composed of glass a piezoelectric material. The Bernoulli's beam theory is adopted for mechanical deformations, and the electric potential field of the piezoelectric material is assumed such that the divergence-free requirement of the electrical displacements is satisfied. The accuracy of the analytic model is assessed by comparing the resonance frequencies obtained by the analytic model with those obtained by the finite element method. The model developed can be used as a tool for designing piezoelectric actuators such as micro-pumps.

## 1. INTRODUCTION

Because piezoelectric materials are widely used as actuators and sensors in the form of composites, analyses of such composite structures have attracted much attention. Examples include the analytical modeling of a beam with surface-bonded or embedded piezoelectric sensors and actuators (Bailey and Hubbard [1], Lee [2], Wang and Rogers [3]), use of piezoelectric materials in composite laminates and for vibration control (Wang et al. [4], Blanguernon et al. [5]). The use of the finite element method in the analysis of piezoelectric coupled structures has been studied (Robinson and Reddy [6], Huang and Park [7], Saravanos and Heyliger [8], Kim et al. [9]) and executed in commercial FEA codes (HKS Inc [10]).

Crawley and de Luis [11] developed a uniform strain model with surface bonded and embedded piezoelectric actuator patches. Models for composite structures with piezoelectric materials as sensors and actuators have also been published (Han and Lee [12]). Kunkel et al. [13] and Kocbach et al. [14] have studied the natural vibrational modes of axially symmetric piezoelectric ceramic disks by the finite element method. Q Wang et al. [15] have considered the free vibration analysis of a piezoelectric coupled circular plate. They used the Kirchhoff thin plate model for the displacement field and assumed a quadratic variation for the electrical potential in the thickness direction..

Most of the aforementioned works are concerned with symmetric structures with flexural deformation only. In this paper an asymmetric beam, in which both flexural and stretching deformations occur simultaneously, is considered. The beam is composed of a glass with a piezoelectric material poled in the thickness direction. The Bernoulli's beam theory is adopted for mechanical deformations and the same assumption on the electric potential adopted in Q Wang et al. [15] is followed. The governing equations are derived by the principle of virtual work. The validation of the proposed model is done by comparing the results from the model and those obtained by the finite element method.

## 2. BASIC EQUATIONS

Under the so-called quasi-static approximation, the electric field is assumed to be irrotational and the Maxwell equations can be simplified as

$$\nabla \cdot \mathbf{D} = 0, \quad (1)$$
$$\nabla \times \mathbf{E} = 0, \quad (2)$$

where $\mathbf{D}$ and $\mathbf{E}$ denote as the electric displacement and electric field intensity, respectively. From Eq. (2), the electric field can be expressed in term of the electric potential $\Phi$ as

$$\mathbf{E} = -\nabla \Phi. \quad (3)$$

The constitutive equations for a piezoelectric material can be described as

$$\begin{aligned}\sigma_\alpha &= c_{\alpha\beta}\gamma_\beta - e_{i\alpha}E_i \\ D_i &= \varepsilon^S_{ij}E_j + e_{i\alpha}\gamma_\alpha\end{aligned}, \quad (4)$$

where $e_{i\alpha}$ is the piezoelectric stress constants and $c_{\alpha\beta}$ is the elastic stiffness at constant $\mathbf{E}$.

The governing equations for the electrical and mechanical fields may be derived by the principle of

---

[1] Research Assistant  * Corresponding Author
[2] Post-Doctoral Research Associate
[3] Professor

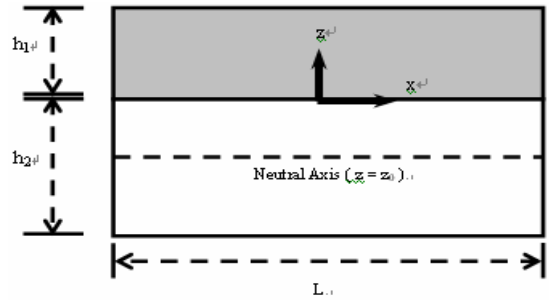

Figure 1. Configuration of the asymmetric piezoelectric composite beam.

virtual work given as
$$\delta W = 0, \quad (5)$$
where $\delta W$ stands for the virtual work done by all forces acting on a system that undergoes an infinitesimal virtual displacement $\delta u_i$. The virtual displacement $\delta u_i$ must be compatible with the specified displacement boundary conditions. The applied forces can be classified into three parts: the external forces $F_{ext}$, the internal forces $F_{int}$, and the inertial forces $F_{ine}$. The virtual work $\delta W$ may be expressed as
$$\delta W = \delta W_{ext} + \delta W_{int} + \delta W_{ine} = 0, \quad (6)$$
where $\delta W_{ext}$ is the virtual work of external forces $F_{ext}$; $\delta W_{int}$ is the virtual work of internal forces $F_{int}$; $\delta W_{ine}$ is the virtual work of inertial forces $F_{ine}$.
From the definitions of $\delta W_{ext}$, $\delta W_{int}$, and $\delta W_{ine}$, Eq. (6) can be derived as
$$\iiint_R \left( f_i \delta u_i - \sigma_{ij} \delta \varepsilon_{ij} - \rho \ddot{u}_i \delta u_i \right) dV + \iint_S \overline{t}_i \delta u_i dS = 0, \quad (7)$$
$$\iiint_R \left( \rho_e \delta \phi - D_i \delta E_i \right) dV + \iint_S \overline{D}_n \delta \phi dS = 0. \quad (8)$$

Figure 1 shows a glass beam with a piezoelectric beam mounted on its surface. The length of this model is expressed as $L$; the thickness of the piezoelectric material as $h_1$; the thickness of the glass as $h_2$; the total thickness of this composite beam model as $h$.
According to Bernoulli's beam theory, the displacement field is assumed as follows:
$$\begin{cases} u_1 = u(x,t) - (z - z_0)\dfrac{\partial w}{\partial x} \\ u_3 = w(x,t) \end{cases} \quad -h_2 \le z \le h_1, \quad (9)$$
where $u_1$ and $u_3$ are the components of displacements in the $x-$ and $z-$ directions of the plane, respectively. The strain $\gamma_1 = u_{1,1}$ is given by
$$\gamma_1 = u_{1,1} = u' - (z - z_0)w'', \quad (10)$$
where $(f)'$ is the partial derivative of the function $(f)$ with respect to the spatial coordinate $x$.

For the electric potential, a quadratic variation of the electrical potential in the $z-$ direction is assumed as (Wang et al.,2001)
$$\phi(x,z,t) = z(h_1 - z)\varphi(x,t) \quad 0 \le z \le h_1, \quad (11)$$
where $z$ is measured from the interface, $h_1$ is the thickness of the piezoelectric beam, and $\varphi(x,t)$ is a function of $x$ and $t$. From equation (11) and the constitutive equations for the piezoelectric solids, the components of the electric field **E** and electric displacement **D** are obtained as follows:
$$E_1 = -\frac{\partial \phi}{\partial x} = z(z - h_1)\varphi', \quad (12)$$
$$E_3 = -\frac{\partial \phi}{\partial z} = (2z - h_1)\varphi, \quad (13)$$
$$D_1 = \overline{\varepsilon}_{11} E_1 = \overline{\varepsilon}_{11} z(z - h_1)\varphi', \quad (14)$$
$$D_3 = \overline{e}_{31}\gamma_1 + \overline{\varepsilon}_{33} E_3 = \overline{e}_{31}\left[u' - (z - z_0)w''\right] + \overline{\varepsilon}_{33}(2z - h_1)\varphi, \quad (15)$$
where $\overline{\varepsilon}_{11} = \varepsilon_{11}$ and $\overline{\varepsilon}_{33} = \varepsilon_{33} + (e_{33}^2 / c_{33})$ are reduced dielectric constants of the piezoelectric material assuming $\gamma_2 = 0$. With Eqs. (9) and (11), Eqs. (7) and (8), respective, become.
$$\int_{-h_2}^{h_1} \left( \sigma_{11}\delta u_{1,1} + \rho \ddot{u}_1 \delta u_1 + \rho \ddot{u}_3 \delta u_3 \right) dz dx = 0, \quad (16)$$
and
$$\int_{-h_2}^{h_1} \left( D_1 \delta E_1 + D_3 \delta E_3 \right) dz dx = 0. \quad (17)$$
Equations (16) and (17) to be further expressed as
$$\int_{-h_2}^{h_1} \left\{ \left[ -N_1' + (\rho_0 \ddot{u} - \rho_1 \ddot{w}') \right] \delta u \right. $$
$$\left. + \left[ -M_1'' + (\rho_1 \ddot{u}' - \rho_2 \ddot{w}'') + \rho_0 \ddot{w} \right] \delta w \right\} dx = 0, \quad (18)$$
$$\int_0^{h_1} \left( -\mathscr{D}_1' + \mathscr{D}_3 \right) \delta \varphi dx = 0, \quad (19)$$
where $\rho_k = \int_{-h_2}^{h_1} \rho (z - z_0)^k dz$,
$$\mathscr{D}_1 = \int_0^{h_1} D_1 z(z - h_1) dz = \overline{\varepsilon}_{11} \frac{h_1^5}{30} \varphi'$$
$$\mathscr{D}_3 = \int_0^{h_1} D_3 (2z - h_1) dz = Fw'' + \overline{\varepsilon}_{33} \frac{h_1^3}{3} \varphi,$$

and $N_1$ and $M_1$ represent the $x$ component of the resultant force and moment, respectively.

For Eqs. (18) and (19) to be satisfied for arbitrary $\delta u$, $\delta w$, and $\delta \varphi$, we must have

$$-N_1' + (\rho_0 \ddot{u} - \rho_1 \ddot{w}') = 0, \quad (20)$$

$$-M_1'' + (\rho_1 \ddot{u}' - \rho_2 \ddot{w}'') + \rho_0 \ddot{w} = 0, \quad (21)$$

$$-\mathscr{D}_1' + \mathscr{D}_3 = 0. \quad (22)$$

The stress component $\sigma_{11}^{(1)}$ in the piezoelectric plane is expressed as

$$\sigma_{11}^{(1)} = \bar{c}_{11}^{(1)} \gamma_1 - \bar{e}_{31} E_3, \quad (23)$$

and the stress component $\sigma_{11}^{(2)}$ in the glass plane is expressed as

$$\sigma_{11}^{(2)} = \bar{c}_{11}^{(2)} \gamma_1. \quad (24)$$

$\bar{c}_{11}^{(1)}$, $\bar{e}_{31}$, and $\bar{c}_{11}^{(2)}$ are transformed reduced material constants of the piezoelectric material and glass, respectively, assuming $\gamma_2 = 0$ and $\sigma_3 = 0$. $\bar{c}_{11}^{(1)}$, $\bar{e}_{31}$, and $\bar{c}_{11}^{(2)}$ are given by

$$\bar{c}_{11}^{(1)} = c_{11}^{(1)} - \frac{(c_{13}^{(1)})^2}{c_{33}^{(1)}} \quad \bar{c}_{11}^{(2)} = c_{11}^{(2)} - \frac{(c_{13}^{(2)})^2}{c_{33}^{(2)}} \quad \bar{e}_{31} = e_{31} - \frac{c_{13}^{(1)}}{c_{33}^{(1)}} e_{33},$$

where $c_{11}^{(1)}$ and $c_{11}^{(2)}$ are the elastic moduli of the piezoelectric material and the glass material in the $x$-direction, respectively.

The resultant force and moment can be expressed, respectively, as

$$N_1 = A_{11} u' - B_{11} w'', \quad (25)$$

$$M_1 = B_{11} u' - D_{11} w'' + F\varphi, \quad (26)$$

where

$$A_{11} = \int_{-h_2}^{h_1} \bar{c}_{11} dz = \bar{c}_{11}^{(1)} h_1 + \bar{c}_{11}^{(2)} h_2,$$

$$B_{11} = \int_{-h_2}^{h_1} \bar{c}_{11} (z - z_0) dz$$
$$= \left[ \left( \bar{c}_{11}^{(1)} \frac{h_1^2}{2} - \bar{c}_{11}^{(2)} \frac{h_2^2}{2} \right) - z_0 \left( \bar{c}_{11}^{(1)} h_1 + \bar{c}_{11}^{(2)} h_2 \right) \right],$$

$$D_{11} = \int_{-h_2}^{h_1} \bar{c}_{11} (z - z_0)^2 dz$$
$$= \bar{c}_{11}^{(1)} \left( \frac{h_1^3}{3} - h_1^2 z_0 + h_1 z_0^2 \right) + \bar{c}_{11}^{(2)} \left( \frac{h_2^3}{3} + h_2^2 z_0 + h_2 z_0^2 \right),$$

$$F = \int_0^{h_1} -\bar{e}_{31} (2z - h_1)(z - z_0) dz = -\frac{\bar{e}_{31}}{6} h_1^3.$$

Let $z_0$ be chosen such that $B_{11} = 0$, namely

$$z_0 = \frac{1}{2} \frac{\bar{c}_{11}^{(1)} h_1 - \bar{c}_{11}^{(2)} h_2}{\bar{c}_{11}^{(1)} h_1 + \bar{c}_{11}^{(2)} h_2}. \quad (27)$$

Furthermore, substituting Eqs. (25) and (26) into Eqs. (20) and (21), respectively, yields two decoupled equations. Moreover, Eq. (22) also can be expressed by $w$ and $\varphi$ as

$$A_{11} u'' = \rho_0 \ddot{u}, \quad (28)$$

$$-D_{11} w^{(4)} + F\varphi'' = -\rho_2 \ddot{w}'' + \rho_0 \ddot{w}, \quad (29)$$

$$\eta_1 \varphi'' + \varphi + \eta_2 w'' = 0, \quad (30)$$

where $\eta_1 = -\frac{h_1^2}{10} \frac{\bar{\varepsilon}_{11}}{\bar{\varepsilon}_{33}}$, and $\eta_2 = \frac{3F}{h_1^3 \bar{\varepsilon}_{33}}$.

Solving Eqs. (29) and (30) for $\varphi$ gives

$$\varphi(x,t) = -\eta_1 \frac{1}{F} \left[ D_{11} w^{(4)} - \rho_2 \ddot{w}'' + \rho_0 \ddot{w} \right] - \eta_2 w''. \quad (31)$$

Differentiating the above equation with respect to the variable $x$ twice gives

$$\varphi'' = -\eta_1 \frac{1}{F} \left[ D_{11} w^{(6)} - \rho_2 \ddot{w}^{(4)} + \rho_0 \ddot{w}'' \right] - \eta_2 w^{(4)}. \quad (32)$$

Substituting Eq. (32) into Eq. (29) gives a decoupled sixth-order differential equation for $w$, namely

$$\eta_1 \left[ D_{11} w^{(6)} - \rho_2 \ddot{w}^{(4)} + \rho_0 \ddot{w}'' \right] - \rho_2 \ddot{w}'' + \bar{D} w^{(4)} + \rho_0 \ddot{w} = 0, \quad (33)$$

where $\bar{D} = D_{11} + \eta_2$.

## 3. FREE VIBRATION

Let $w(x,t) = \hat{w}(x) e^{i\omega t}$, where $\hat{w}(x)$ is the amplitude of the $z$-direction displacement as a function of $x$ only. Rewriting equation (33) in terms of $\hat{w}(x)$ and eliminating the term $e^{i\omega t}$ gives a sixth-order differential equation, namely

$$\eta_1 D_{11} \hat{w}^{(6)}(x) + \left( \eta_1 \rho_2 \omega^2 + \bar{D} \right) \hat{w}^{(4)}(x) + \left( \rho_2 - \eta_1 \rho_0 \right) \omega^2 \hat{w}''(x) - \rho_0 \omega^2 \hat{w}(x) = 0 \quad (34)$$

In general $\eta_1$ and $\rho_2$ in Eq. (34) are much smaller than the other parameters. The terms containing $\eta_1$ and $\rho_2$ can thus be neglected except the term $(\rho_2 - \eta_1 \rho_0) \omega^2$ and Eq. (34) becomes

$$\hat{w}^{(4)}(x) + 2\alpha^2 \hat{w}''(x) - \beta^4 \hat{w}(x) = 0, \quad (35)$$

where $\alpha = \left[ \frac{(\rho_2 - \eta_1 \rho_0) \omega^2}{2\bar{D}} \right]^{\frac{1}{2}}$, and $\beta = \left( \frac{\rho_0 \omega^2}{\bar{D}} \right)^{\frac{1}{4}}$.

The solution to equation (35) can be written as
$$\hat{w}(x) = b_1 \cosh(n_1 x) + b_2 \sinh(n_2 x) + b_3 \cos(n_3 x) + b_4 \sin(n_4 x), \quad (36)$$

where $n_{1,2} = \pm \sqrt{-\alpha^2 + \sqrt{\alpha^4 + \beta^4}}$,

$n_{3,4} = \pm \sqrt{-\alpha^2 - \sqrt{\alpha^4 + \beta^4}}$,

and $b_i$ ($i = 1 \sim 4$) are constants.

Since $\hat{w}(x)$ is symmetric with respect to $x$, the constants $b_2$ and $b_4$, must vanish. Then the solution reduces to

$$\hat{w} = b_1 \cosh(n_1 x) + b_3 \cos(n_3 x), \qquad (37)$$

Table 1. Material properties of the asymmetric piezoelectric composite beam.

|  | Glass | Piezoelectric material (PZT-5A) | |
|---|---|---|---|
| Elastic properties ($N\ m^{-2}$) | $c_{11} = 16.5 \times 10^{10}$ | $c_{11}^E = 12.1 \times 10^{10}$ | $c_{33}^E = 11.1 \times 10^{10}$ |
|  | $c_{12} = 6.3 \times 10^{10}$ | $c_{12}^E = 7.54 \times 10^{10}$ | $c_{44}^E = 2.11 \times 10^{10}$ |
|  | $c_{13} = 6.3 \times 10^{10}$ | $c_{13}^E = 7.52 \times 10^{10}$ | $c_{66}^E = 2.26 \times 10^{10}$ |
| Mass density ($kg\ m^{-3}$) | 7750 | 2330 | |
| $e_{31}(C\ m^{-2})$ |  | -5.4 | |
| $e_{33}(C\ m^{-2})$ |  | 15.8 | |
| $e_{15}(C\ m^{-2})$ |  | 12.3 | |
| $\varepsilon_{11}^S(F\ m^{-1})$ |  | $8.110264 \times 10^{-9}$ | |
| $\varepsilon_{33}^S(F\ m^{-1})$ |  | $7.34882 \times 10^{-9}$ | |

Table 2. Comparison of the first three resonance frequencies for the modified model.

| Mode no | Analytical results | FEA results | Error (%) |
|---|---|---|---|
| 1 | 4.52E+04 | 4.48E+04 | 0.87 |
| 2 | 3.81E+05 | 3.60E+05 | 5.56 |
| 3 | 9.50E+05 | 8.57E+05 | 9.72 |

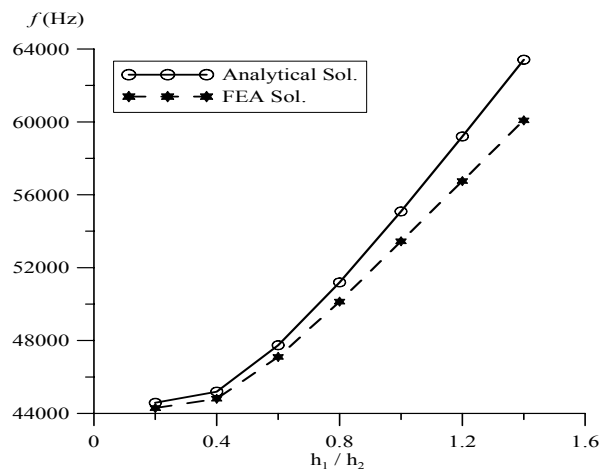

Figure 2. Variations of the first resonance the frequency with different thickness ratios.

In this paper the beam is assumed to be simply supported with the boundary conditions given by

$$\hat{w}(l/2) = \hat{M}_1(l/2) = 0, \qquad (38)$$

and the resulting characteristic equation can be shown to be:

$$\cos\left(n_3 \frac{l}{2}\right) = 0. \qquad (39)$$

The frequency is obtained as

$$\omega = \left(\frac{m\pi}{l}\right)^2 \sqrt{\frac{\overline{D}}{\rho_0 + (m\pi/l)^2 (\rho_2 - \eta_1 \rho_0)}}, \qquad (40)$$

$m = 1, 3, 5, \cdots$ (odd integers)

## 4. RESULTS

The resonant frequencies for the simply-supported beam were calculated using Eq. (40). The material parameters for the beam structure used in the numerical calculators are listed in table 1. Table 2 lists the first three resonance frequencies of this piezoelectric composite beam structure using Eq. (40) and finite element simulations, respectively. In this example, the thickness of the piezoelectric material is 200μm and that of the glass is 500μm. In addition, the variations of the first resonance frequency with the thickness ratio of the piezoelectric material and glass from 0.2 to 1.4 are displayed in figure 2. It can be seen that the results by Eq. (40) agree closely with those obtained from the finite element analysis.

## 5. CONCLUSIONS

In this paper the resonant frequency of an asymmetric piezoelectric composite beam is studied. An analytic expression for the resonant frequency has been derived. The resonant frequencies predicted by the analytic solution are shown to be close to those computed by finite element simulations.

## 6. ACKNOWLEDGEMENT

The research was supported by the National Science Council of Taiwan under grant NSC 95-2221-E-002-228 and partially supported by the Ministry of Economic Affair under grant no 95-EC-17-A-05-S1-017 (WHAM-BioS). Computational support was provided by National Center for High-performance Computing.

## 7. REFERENCES

[1] T. Bailey and J.E. Hubbard, "Distributed piezoelectric-polymer active vibration control of a cantilever beam," *Journal of Guidance*, Vol.8, 1985, pp. 605-611.

[2] C.-K. Lee, "Theory of laminated piezoelectric plates for the design of distributed sensors/actuators. Part I：governing equations and reciprocal relationships," *Journal of the Acoustic Society of America*, Vol.87, 1990, pp. 1144-1158.

[3] B.-T. Wang and C.A. Rogers, "Laminate plate theory for spatially distributed induced strain actuators," *Journal of Computers and Materials*, Vol.25, 1991, pp. 433-452.

[4] Q.Wang, K.M., Liew, and D.J. Wang, "Some issues of control of structures using piezoelectric actuators," *Proc. SPIE*, Vol. 2921, 1997, pp. 425-430

[5] A. Blanguernon, F. Lene, and M. Bernadou, "Active control of a beam using a piezoceramic element," *Smart Mater. Struct.*, Vol.8, 1999, pp. 116-124.

[6] D.H. Robinson, and J.N. Reddy, "Analysis of piezoelectrically actuated beams using a layer-wise displacement theory," *Computer and Structures*, Vol.41, 1991, pp. 265-279.

[7] W.-S. Huang and H.C. Park, "Finite element modeling of piezoelectric sensors and actuators," *AIAA Journal*, Vol. 31, 1993, pp. 930-937.

[8] D.A. Saravanos and P.R. Heyliger, "Coupled layer-wise analysis of composite beams with embedded piezoelectric sensors and actuators," *J. Intell. Mater. Syst. Struct.*, Vol.6, 1995, pp. 350-363.

[9] J. Kim, V.V. Varadan, V.K., Varadan, and X.Q. Bao, "Finite element modeling of a smart cantilever plate and comparison with experiments," *Smart Mater. Struct.*, Vol. 5, 1996, pp. 165-170.

[10] HKS Inc *ABAQUS User's Manual* (*version 5.2*). Providence, RI: Hibbitt, Karlsson and Sorenson Inc, 1993.

[11] E.F. Crawley and J. de Luis, "Use of piezoelectric actuators as elements of intelligent structures," *AIAA Journal*, Vol.25, 1987, pp. 1373-1385.

[12] J.H. Han and L. Lee, "Analysis of composite plates with piezoelectric actuators for vibration control using layerwise displacement theory," *Composites* B, Vol.29, 1998, pp. 621-632.

[13] H.A. Kunkel, S. Locke, and B. Pikeroen "Finite-element analysis of vibrational modes in piezoelectric ceramic disks," *IEEE Transactions on ultrasonics, ferroelectrics, and frequency control*., Vol.37, no.4., 1990, pp. 316-328.

[14] J. Kocbach, P. Lunde, and M. Vestrheim, "Resonance frequency spectra with convergence tests of piezoceramic disks using the finite element method," *ACUSTICA*, Vol.87, no.2, 2001, pp. 271 – 285.

[15] Q. Wang, S.T. Quek, C.T. Sun, and X. Liu, "Analysis of piezoelectric coupled circular plate," *Smart Mater. Struct.* Vol.10, 2001, pp. 229-239.